\documentclass[%
 reprint,
superscriptaddress,
 amsmath,amssymb,
 aps,
]{revtex4-2}
\usepackage{amssymb}
\usepackage{graphicx}
\usepackage{dcolumn}
\usepackage{bm}
\usepackage[colorlinks, linkcolor=blue, citecolor=blue, urlcolor=blue]{hyperref}

\begin{document}

\preprint{APS/123-QED}

\title{{Interpretable descriptors enable prediction of hydrogen-based superconductors at moderate pressures}}

\author{Jiawei Chen}
\affiliation{School of Physics and Optoelectronic Engineering, Guangdong University of Technology, Guangzhou, Guangdong 510006, China}

\author{Junhao Peng}
\email{junhaopeng@outlook.com}
\affiliation{School of Physics and Optoelectronic Engineering, Guangdong University of Technology, Guangzhou, Guangdong 510006, China}
\affiliation{Guangdong Provincial Key Laboratory of Sensing Physics and System Integration Applications, Guangdong University of Technology, Guangzhou, Guangdong 510006, China}

\author{Yanwei Liang}
\affiliation{School of Physics and Optoelectronic Engineering, Guangdong University of Technology, Guangzhou, Guangdong 510006, China}

\author{Renhai Wang}
\affiliation{School of Physics and Optoelectronic Engineering, Guangdong University of Technology, Guangzhou, Guangdong 510006, China}

\author{Huafeng Dong}
\email{hfdong@gdut.edu.cn}
\affiliation{School of Physics and Optoelectronic Engineering, Guangdong University of Technology, Guangzhou, Guangdong 510006, China}
\affiliation{Guangdong Provincial Key Laboratory of Sensing Physics and System Integration Applications, Guangdong University of Technology, Guangzhou, Guangdong 510006, China}

\author{Wei Zhang}
\affiliation{School of Physics and Optoelectronic Engineering, Guangdong University of Technology, Guangzhou, Guangdong 510006, China}
\affiliation{Guangdong Provincial Key Laboratory of Sensing Physics and System Integration Applications, Guangdong University of Technology, Guangzhou, Guangdong 510006, China}

\date{\today}

\begin{abstract}
Room-temperature superconductivity remains elusive, and hydrogen-base compounds—despite remarkable transition temperatures($T_\mathrm{c}$)—typically require extreme pressures that hinder application. To accelerate discovery under moderate pressures, an interpretable framework based on symbolic regression is developed to predict $T_\mathrm{c}$ in hydrogen-based superconductors. A key descriptor is an integrated density of states (IDOS) within $\pm$ 1 eV of the Fermi level ($E_\mathrm{F}$), which exhibits greater robustness than conventional single-point DOS features. The resulting analytic model links electronic-structure characteristics to superconducting performance, achieves high accuracy ($\mathrm{RMSE}_{\mathrm{train}}$ = 20.15 K), and generalizes well to external datasets. By relying solely on electronic-structure calculations, the approach greatly accelerates materials screening. Guided by this model, four hydrogen-based candidates are identified and validated via calculation: Na$_{2}$GaCuH$_{6}$ with $T_\mathrm{c}$ = 42.04 K at ambient pressure (exceeding MgB$_{2}$), and NaCaH$_{12}$, NaSrH$_{12}$, and KSrH$_{12}$ with $T_\mathrm{c}$ up to 162.35 K, 86.32 K, and 55.13 K at 100 GPa, 25 GPa, and 25 GPa, respectively. Beyond rapid screening, the interpretable form clarifies how hydrogen-projected electronic weight near $E_\mathrm{F}$ and related features govern $T_\mathrm{c}$ in hydrides, offering a mechanism-aware route to stabilize high-$T_\mathrm{c}$ phases at reduced pressures.
\end{abstract}

\maketitle


\section{Introduction}

Hydrogen-base compounds provide a natural setting for conventional superconductivity, where light hydrogen modes and robust covalent frameworks jointly boost electron–phonon coupling and phonon energy scales, pushing superconducting transition temperatures ($T_\mathrm{c}$) toward ambient temperatures under compression \cite{wignerPossibilityMetallicModification1935, ashcroftHydrogenDominantMetallic2004,
duan2017structure}. Within such frameworks, chemical “precompression” \cite{ashcroftHydrogenDominantMetallic2004} embedded in the lattice lowers the metallization threshold of extended hydrogen networks, enabling superconducting phases at substantially reduced external pressure. Clathrate-like hydrides exemplify this design principle, achieving $T_\mathrm{c}$ values on the order of 200 - 250 K at megabar pressures, thereby positioning hydrogen-base materials as leading candidates for room-temperature superconductivity and underscoring the predictive power of computation-guided materials discovery\cite{wang2012superconductive,ma2022hightemperature,troyanAnomalousHighTemperatureSuperconductivity2021b,songHighT_rmSuperconductivity2021,duan2014pressureinduced,duan2015pressureinduced,drozdovConventionalSuperconductivity2032015c,liu2017potential,drozdovSuperconductivity250Lanthanum2019d}.

Despite these advances, practical deployment is obstructed by the requirement of ultrahigh pressures for structural stability. Hydrides that retain high-$T_\mathrm{c}$ while remaining stable at moderate-to-low pressures are therefore a central target. Conventional first-principles workflows accurately capture electron–phonon interactions but are computationally prohibitive for exhaustive exploration across vast chemical spaces, creating a discovery bottleneck precisely where rapid iteration is most needed\cite{pickard2020superconducting}.

Data-driven approaches offer a route to overcome this bottleneck while preserving physical insight. Instead of opaque black-box predictors\cite{zhong2022explainable}, symbolic-regression models \cite{xieMachineLearningSuperconducting2022} can deliver compact, interpretable formulas that connect $T_\mathrm{c}$ to physically motivated descriptors—such as electron-localization measures, molecular/structural indices, hydrogen-projected density of states at the Fermi level ($E_\mathrm{F}$) \cite{belliStrongCorrelationElectronic2021a,dimauroMolecularityFastEfficient2024,novoaTcESTIMEPredictingHightemperature2024,winesDatadrivenDesignHigh2024}, and proxies for EPC strength and characteristic phonon scales. By revealing quantitative structure–property relations, such models enable fast prescreening and mechanism-aware optimization under pressure constraints. In this Letter, a physically constrained symbolic-regression framework is developed for hydrogen-base superconductors, producing an analytic $T_\mathrm{c}$ descriptor from readily computed features, validating its predictivity across known hydrides, and identifying chemically plausible candidates predicted to retain high-$T_\mathrm{c}$ under moderate pressures.

\section{Computational Methods}

All electronic structure calculations were performed within the framework of density functional theory (DFT) by the Vienna \textit{Ab initio} Simulation Package (VASP) \cite{PhysRevB.54.11169,kresseEfficiencyAbinitioTotal1996}. The exchange--correlation interactions among electrons were described using the Perdew--Burke--Ernzerhof (PBE) functional under the generalized gradient approximation (GGA) \cite{perdewGeneralizedGradientApproximation1996}. To ensure reliable results, especially under high pressures, a plane-wave cutoff energy of 1000~\text{eV} was employed. Brillouin zone integrations were performed using a Monkhorst--Pack \textit{k}-point mesh with a reciprocal-space resolution of $2\pi \times 0.03~\text{\AA}^{-1}$.

Phonon dispersions and the electron--phonon coupling (EPC) parameter $\lambda$ were computed with the Quantum ESPRESSO (QE) \cite{giannozzi2020quantum, giannozzi2017advanced, giannozzi2009quantum} package using optimized norm-conserving Vanderbilt (ONCV) \cite{hamann2013optimized} pseudopotentials. The plane-wave kinetic-energy cutoff was 80 Ry. Self-consistent DFT calculations employed a 16$\times$16$\times$16 Monkhorst--Pack k-point grid \cite{monkhorst1976special}. Density Functional Perturbation Theory (DFPT) \cite{baroni2001phonons} calculations used a 2$\times$2$\times$2 phonon $q$ grid.

For the Brillouin-zone integrations entering the electron--phonon coupling (EPC) calculation, the reported $T_\mathrm{c}$ values correspond to a double-delta broadening of $\sigma = 0.05$. Based on the Eliashberg spectral function $\alpha^2F(\omega)$, the electron--phonon coupling constant $\lambda$ and the logarithmic-average phonon frequency $\omega_{\mathrm{log}}$ were evaluated. The $T_\mathrm{c}$ was then estimated using the Allen--Dynes modified McMillan formula \cite{PhysRevB.6.2577, PhysRevB.12.905} with a Coulomb pseudopotential of $\mu^\ast = 0.10$.

\section{Result and dicussion}

\begin{figure}[t]
  \centering
  \includegraphics[width=\linewidth]{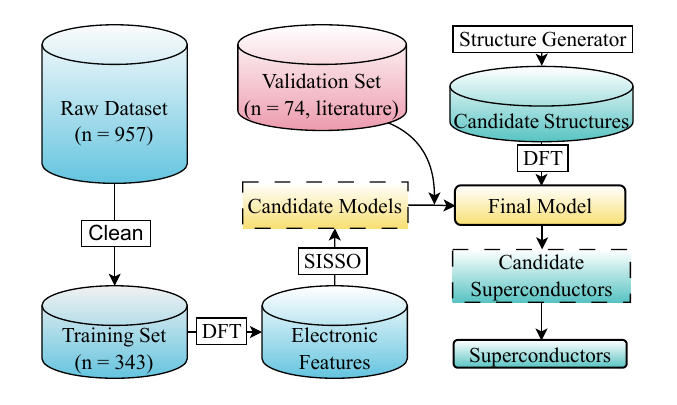}
  \caption{Workflow for constructing and applying a prediction model of $T_\mathrm{c}$ in hydrogen-based superconductors using first-principles calculations and symbolic regression.}
  \label{fig:fig1}
\end{figure}

The workflow for constructing and applying our $T_\mathrm{c}$ prediction model is illustrated in Fig.\ref{fig:fig1}. We began with a database of hydrogen-based superconductors obtained from first-principles calculations, comprising 957 samples \cite{winesDatadrivenDesignHigh2024}. Data were cleaned based on two criteria: (1) exclusion of structures with imaginary phonon frequencies to guarantee dynamical stability; and (2) retention of only converged calculations yielding transition temperatures within the physically reasonable range of 0--500 K. This filtering process resulted in a training set of 343 samples.  Electronic structure features were extracted using VASP and used as input variables for the sure independence screening and sparsifying operator (SISSO) \cite{ouyang2018sisso}. To enhance the generalization capability of the model, we incorporated an external validation set comprising 72 samples collected from published literature \cite{zhangDesignPrinciplesHighTemperature2022,doluiFeasibleRouteHighTemperature2024,cerqueiraSearchingMaterialsSpace2024, wang2025prediction}. This validation set was used to evaluate candidate models and guide the selection of the final prediction model. In the application stage, we employed a structure generator to produce large sets of hydride structures, applied the final prediction model to identify potential superconductor candidates. Subsequently, we carried out electron--phonon coupling calculations using QE to verify the candidates, ultimately identifying hydrogen-based superconductors with promisingly high transition temperatures.

\begin{figure*}[t]
    \centering
    \includegraphics[width=0.8\textwidth]{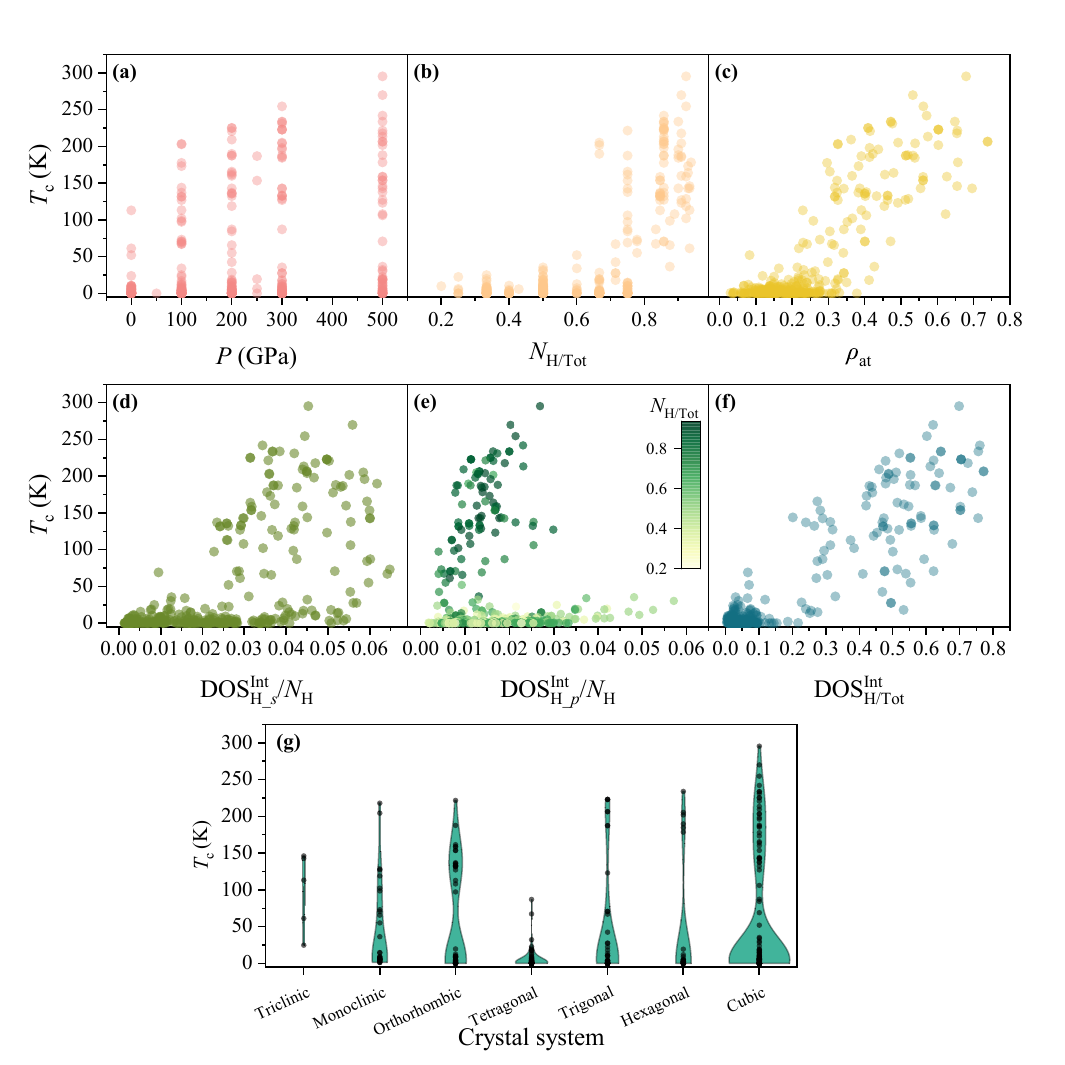}
    \caption{Relationships between key parameters and $T_\mathrm{c}$ in the training set. (a) Pressure ($P$); (b) hydrogen atomic fraction (${N_\text{H/Tot}}$); (c) atomic number density ($\rho_{\text{at}}$); (d) integrated hydrogen $s$-orbital partial density of states (PDOS) normalized by hydrogen atom count (${\mathrm{DOS}^{\text{Int}}_{\text{H}\_s}}$); (e) integrated hydrogen p-orbital PDOS normalized by hydrogen atom count, color-coded by $N_{\text{H/Tot}}$ ( ${\mathrm{DOS}^{\text{Int}}_{\text{H}\_p}}$), (f) hydrogen-to-total integrated DOS ratio ($\mathrm{DOS}^{\mathrm{Int}}_{\mathrm{H}/\mathrm{Tot}}$); (g) distribution of $T_\mathrm{c}$ across different crystal systems.}
    \label{fig:fig2}
\end{figure*}

Following the workflow outlined in Fig.\ref{fig:fig2} we obtained $T_\mathrm{c}$ and electronic features from 343 training samples after data cleaning. To assess the coverage and representatives of our training set, we conducted statistical analysis on $T_\mathrm{c}$, pressures, hydrogen content, and elemental distributions (Fig. S1). Although imbalances exist (such as higher proportions of low-$T_\mathrm{c}$ samples and hydrogen-rich systems), the dataset covers a broad physico-chemical space, providing a reliable foundation for subsequent model training.

The relationships between key features and $T_\mathrm{c}$ are illustrated in Fig.\ref{fig:fig2}. Pressure analysis (Fig.\ref{fig:fig2}a) reveals that most of high-$T_\mathrm{c}$ superconductors occur in high-pressure regions, consistent with experimental observations of pressure-induced superconductivity where high pressure enhances electron-phonon coupling in hydrogen-based superconductors \cite{duan2017structure}. 

Regarding material composition (Fig.~\ref{fig:fig2}b-c), both hydrogen atomic fraction (\(N_\text{H/Tot}\)) and atomic number density (\(\rho_{\text{at}} = {N_{\text{tot}}}/{V}\)) show positive correlations with \(T_\mathrm{c}\). High-\(T_\mathrm{c}\) emerges predominantly when \(N_\text{H/Tot} >\) 0.65. This phenomenon can be attributed to the high vibrational frequency and light mass of hydrogen atoms, thereby enhancing superconducting pairing strength through the logarithmic average phonon frequency. Atomic number density (Fig.~\ref{fig:fig2}c) displays high correlation with \(T_\mathrm{c}\). This strong correlation can be understood from two perspectives: first, high atomic density implies tighter atomic packing and stronger interatomic interactions, favoring enhanced electron-phonon coupling constant \(\lambda\); second, high density typically corresponds to compressed states under high-pressure conditions, forming an intrinsic connection with the pressure dependence shown in Fig.~\ref{fig:fig2}a. 

As shown in Fig.\ref{fig:fig2}(d-f), we introduce the integrated DOS near the $E_\mathrm{F}$ ($\mathrm{DOS}^\mathrm{Int}$) as a key feature. All calculations perform integration within the energy window of $\pm$ 1 eV, and the rationale for this choice is validated in Fig.S2. Compared to conventional approaches that only consider the DOS at the single point $E_\mathrm{F}$ ($\mathrm{DOS}^{E_\mathrm{F}}$), $\mathrm{DOS}^\mathrm{Int}$  physically represents to the number of activate electrons near the Fermi surface, thereby offering clearer physical meaning. Furthermore, $\mathrm{DOS}^\mathrm{Int}$ exhibits enhanced robustness compared to $\mathrm{DOS}^{E_\mathrm{F}}$ under variations in computational parameters (e.g., k-point sampling, smearing schemes, and pseudopotential choices) as illustrated in Fig.S3(a-d), which can effectively reduce numerical instabilities. Additionally, standard pseudopotentials for H should be used when using this prediction model, as demonstrated in Fig.S3(e-f).

Since $\mathrm{DOS}^\mathrm{Int}$ is an extensive quantity that scales with system size, we performed normalization to convert it into an intensive quantity, enabling meaningful comparisons across different system scales. We normalized by volume or hydrogen atom count to obtain physically meaningful descriptors. 
Fig.\ref{fig:fig2}(d) shows the integrated hydrogen’s $s$-orbital PDOS normalized by hydrogen atom count, which mean the effective $s$-orbital electrons contributed per hydrogen atom (${\mathrm{DOS}^{\text{Int}}_{\text{H}\_s}}$), which exhibits a fan-shaped distribution with $T_\mathrm{c}$, high-$T_\mathrm{c}$ superconductors tend to appear in the region with enhanced $s$-orbital participation, suggesting that hydrogen’s $s$-orbital character is a favorable factor for superconductivity. Fig.\ref{fig:fig2}(e) displays the effective $p$-orbital contributed electrons per hydrogen atom (${\mathrm{DOS}^{\text{Int}}_{\text{H}\_p}}$). It is evident that, in hydrogen-rich compounds, $T_\mathrm{c}$ exhibits a positive correlation with the contribution of hydrogen $p$-orbitals around the Fermi level, suggesting that the emergence of $p$-character in hydrogen plays an important role in enhancing superconductivity. Fig.\ref{fig:fig2}(f) shows the hydrogen-to-total integrated DOS ratio (\(\mathrm{DOS}^{\mathrm{Int}}_{\mathrm{H}/\mathrm{Tot}}\)), representing the ratio of hydrogen effective electrons to total system effective electrons, suggesting a close relationship between the contribution of hydrogen’s effective electrons to the total electron count with superconductivity.

\begin{figure}[t]
    \centering
    \includegraphics[width=\linewidth]{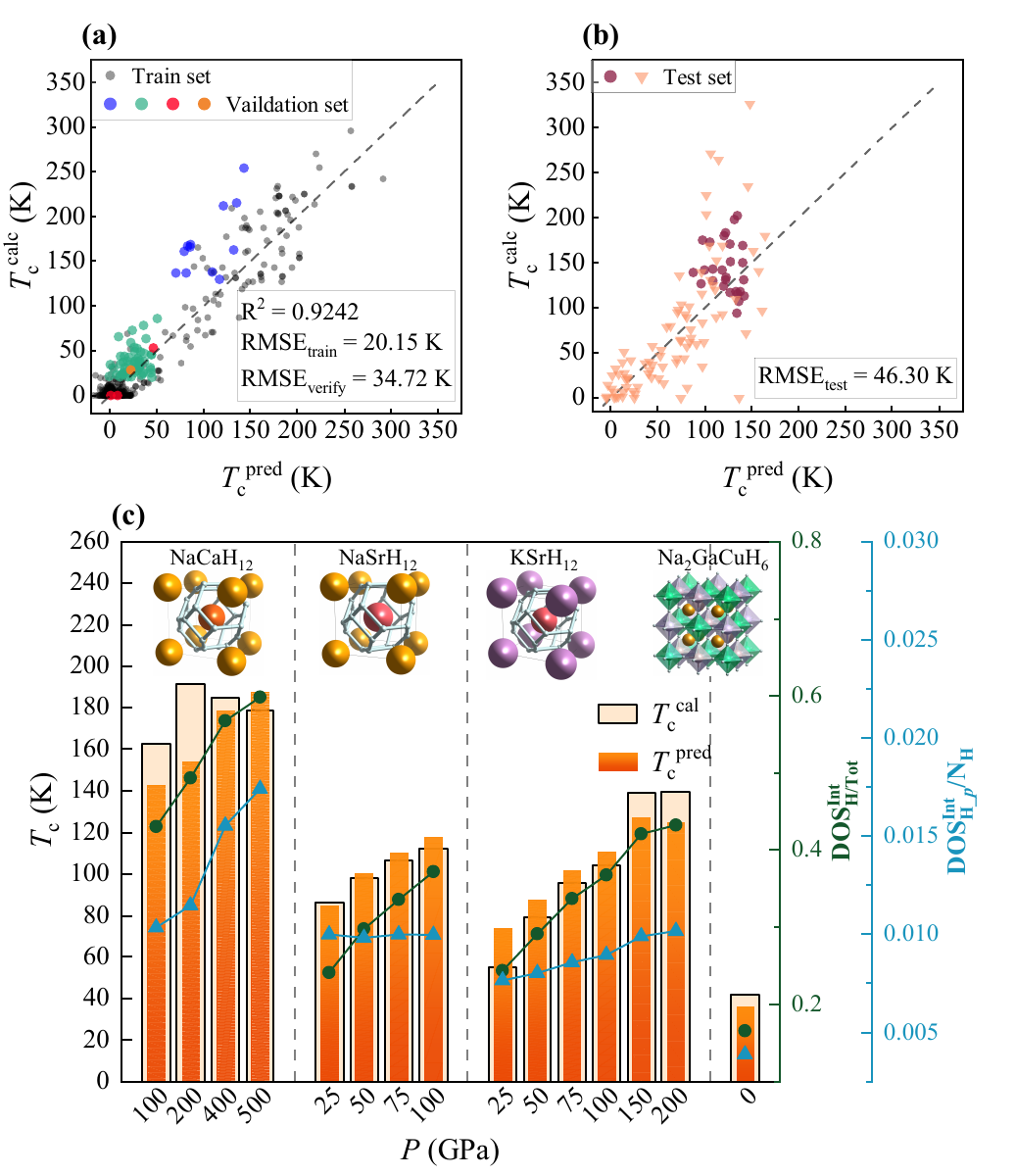}
    \caption{Prediction and application results of the symbolic regression model for the $T_\mathrm{c}$ of hydrogen-based superconductors. (a) Linear fitting of predicted versus calculated values for the training set (gray dots) and the external validation set (colored dots); (b) Linear fitting of $T_\mathrm{c}$ prediction model on the external test set ; (c) The four embedded images represent hydrogen-based superconductors newly found. Data for each superconductor are separated by dashed lines. Light yellow and orange bars indicate the calculated and predicted $T_\mathrm{c}$ of stable structures under different pressures, respectively. The green and blue symbols correspond to the quantities involved in the prediction formula, which are  $\text{DOS}^{\text{Int}}_{\text{H}/\text{Tot}}$ and ${\mathrm{DOS}^{\text{Int}}_{\text{H}\_p}}/{N_\text{H}}$.}
    \label{fig:fig3}
\end{figure}

Fig.\ref{fig:fig2}(g) shows that $T_\mathrm{c}$ distributions vary significantly across different crystal systems. High-$T_\mathrm{c}$ structures are predominantly found in orthorhombic and cubic crystal systems, while other systems contain fewer superconductors with $T_\mathrm{c} >$ 100 K. Notably, the orthorhombic system exhibits a distinct bimodal distribution with peaks in both low-temperature ($<$ 20 K) and high-temperature (100-200 K) regions, a clear gap is observed between them.

Using the SISSO symbolic regression approach, we trained the model on 343 samples, and generated a series of interpretable candidate predictive formulas.
To prevent overfitting on the training set, we constructed an external validation set consisting of 72 hydrogen-based superconductors from recent theoretical studies in the literature. Each candidate prediction model was tested on this validation set, and the root mean square error (RMSE) was evaluated for both training and validation sets. Both errors were jointly used as the selection criterion. The final prediction model for the $T_\mathrm{c}$ of hydrogen-based superconductors can be expressed as:

\begin{equation}
T_\mathrm{c}^{\mathrm{pred}} = a \times H_{act} + b \times D_p + c
\end{equation}

where $H_{act}$ is the hydrogen activity descriptor and $D_p$ is the pressure-weighted dominance, defined as:

\[
H_{act} = \sqrt{\frac{\mathrm{DOS}^{\text{Int}}_{\text{H}\_p}}{N_\text{H}}}\times \text{DOS}^{\text{Int}}_{\text{H}/\text{Tot}},
\]

\[
D_p = \sqrt[3]{\frac{\text{DOS}^{\text{Int}}_\text{H}}{\sum \text{DOS}^{\text{Int}}_\text{X}}} \times P
\]

Here, ${\text{DOS}^{\text{Int}}_\text{H}}/{\sum \text{DOS}^{\text{Int}}_\text{X}}$ is the integrated DOS of non-hydrogen elements as a normalization factor for $\text{DOS}^{\text{Int}}_\text{H}$.

We obtained the coefficients: $a$ = 3930.62 K·(atom/states)$^{1/2}$, $b$ = -0.17 K/GPa, and $c$ = -4.69 K , it is evident that parameter $H_{act}$ provides the dominant contribution to the $T_\mathrm{c}^{\mathrm{pred}}$, while parameter $D_p$ has a much smaller weight, acting more as a fine-tuning term to $T_\mathrm{c}^{\mathrm{pred}}$. 

In the construction of $H_{act}$, the factor $\sqrt{{DOS^{\text{Int}}_{\text{H\_p}}}/{N_\text{H}}}$ erves as a per-atom $p$-orbital activity indicator. By normalizing the integrated $p$-orbital DOS of hydrogen per atom  and taking its square root. When $p$-orbital character emerges at the Fermi surface, it indicates that the electronic system has departed from the simple spherically symmetric ground state and entered a highly active hybridized state with enhanced directionality, spatial extension, and polarizability—essential prerequisites for strong electron-phonon coupling. This activation arises near the $E_\mathrm{F}$ can be through two mechanisms: (i)Under high pressure, the extension and overlap of hydrogen $s$- and $p$-orbitals leads to internal $s$-$p$ hybridization ; and (ii) in hydrides, the hydrogen 1$s$ orbital hybridizes with the $p$-orbitals of neighboring atoms, thereby introducing additional $p$-character into the electronic states. 
The second factor, $\text{DOS}^{\text{Int}}_{\text{H}/\text{Tot}}$, measures the concentration of active electrons of hydrogen near the $E_\mathrm{F}$, is an important descriptor for superconductivity. This descriptor is physically meaningful , because in BCS-type superconductors, the density of states at $E_\text{F}$ directly determines the strength of electron-phonon coupling and thus $T_\mathrm{c}$. A higher $\text{DOS}^{\text{Int}}_{\text{H}/\text{Tot}}$ indicates that hydrogen atoms, with their light mass and high phonon frequencies, dominate the electronic structure near the Fermi surface, creating favorable conditions for strong pairing interactions. 
Only when both factors are large enough does the system possess  high-activity, high-concentration hydrogen electronic states near  the $E_\mathrm{F}$ supporting elevated $T_\mathrm{c}$.

As for parameter $D_p$, it introduces the normalized ratio between hydrogen and non-hydrogen integrated DOS, together with the external pressure. Its role is to capture how the hydrogen contribution is modified under different chemical environments. pressure may introduce competing factors such as bandwidth broadening that slightly suppress $T_\mathrm{c}$. Although its overall effect is weaker than that of $H_\mathrm{act}$, it helps distinguish subtle variations in $T_\mathrm{c}$ among different materials.

The fitting performance of the prediction model is shown in Fig.\ref{fig:fig3}(a). The model yields an RMSE of 20.15 K on the training set and 34.72 K on the validation set. Most data points lie close to the diagonal, indicating strong agreement between predicted and calculated values and confirming the high accuracy of the model ($R^2 = 0.92$).

To further examine the generalization ability, we constructed an external test set from published works \cite{duRoomTemperatureSuperconductivityYb2022,belliStrongCorrelationElectronic2021a}. The test set consists of 141 samples, and the RMSE is 46.30 K, as shown in Fig.\ref{fig:fig3}(b), most points remain close to the diagonal despite these differences, demonstrating that the predictions are consistent with reported values. Overall, this confirms that the symbolic regression model has strong generalization power and can predict $T_\mathrm{c}$ of complex hydrogen-based superconductors beyond the training domain.

\begin{figure}[t]
    \centering
    \includegraphics[width=\linewidth]{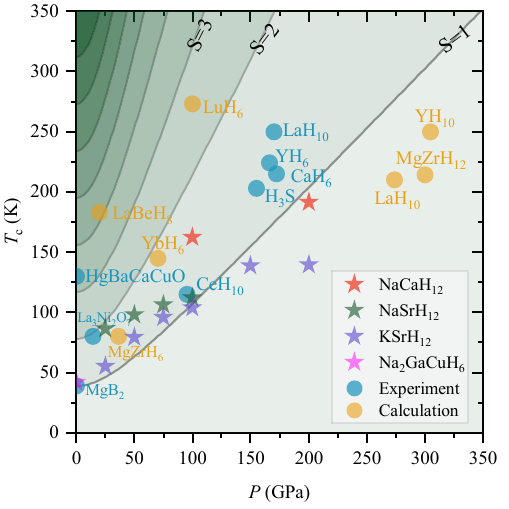}
    \caption{Distribution of $T_\mathrm{c}$ for newly found superconductors under different pressures, as well as known experimental and theoretical superconductors. The background-colored contours represent iso-lines of the superconductivity index S.}
    \label{fig:fig4}
\end{figure}

The prediction model was further applied to identify new hydrogen-based superconductors using two sources of materials: (i) generating candidate structures using PyXtal \cite{fredericksPyXtalPythonLibrary2021} and optimizing them at 100 GPa, and (ii) retrieving “unexplored” structures from InvDesFlow \cite{ouyangHightemperatureSuperconductivity$mathrmLi_2mathrmAuH_6$2025a} that had not been evaluated for superconductivity. Four promising compounds were obtained, NaCaH$_{12}$, NaSrH$_{12}$, KSrH$_{12}$ and Na$_2$GaCuH$_6$. Among them, NaCaH$_{12}$, a derivative of CaH$_6$ \cite{wang2012superconductive,ma2022hightemperature}, becomes dynamically stable at 100 GPa and exhibits $T_\mathrm{c}$ = 162.35 K with a superconductivity index \cite{pickardSuperconductingHydridesPressure2020b} S = 1.51. NaSrH$_{12}$ and KSrH$_{12}$, also CaH$_6$ derivatives, are stable at 25 GPa, with $T_\mathrm{c}$ = 86.32 K (S = 1.86) and 55.13 K (S = 1.25), respectively. Na$_2$GaCuH$_6$ exhibits a $T_\mathrm{c}$ of 42.04 K at ambient pressure, which was not recognized as a promising superconductor in the original study, surpassing the well-known superconductor MgB$_2$ \cite{nagamatsu2001superconductivity}.Notably, this structure was subsequently also identified in the updated version of InvDesFlow \cite{han2025invdesflowalactivelearningbasedworkflow}, where electron-phonon coupling (EPC) calculations confirmed a similar $T_\mathrm{c}$ value of ~42 K, corroborating our prediction and calculation. 
Superconducting performance of the newly identified materials, such as $T_\mathrm{c}$, S and key descriptors for the four compounds across different pressure conditions, as summarized in Table S2.

As shown in Fig.\ref{fig:fig3}(c) and Fig. S4, the predicted and calculated $T_\mathrm{c}$ values agree closely, confirming the reliability of the formula in screening hydrogen-based superconductors. For the newly found materials, the RMSE is 13 K, which indicates a promising prediction performance. Unlike the validation and test sets, we were able to control the computational parameters for the new materials to match those used in the training set. 
As shown in Table S1, the present method achieves comparable accuracy with much higher computational efficiency.
The electronic DOS for each material under different pressures are shown in Fig. S5. Crystal structures with space group information, electronic density of states, phonon dispersion curves, and electron-phonon coupling parameters ($\lambda$ and $\omega_{\mathrm{log}}$) for the four compounds are presented in Supporting Information.

Figure \ref{fig:fig4} illustrates that the newly predicted hydrogen-based superconductors in this study exhibit remarkably high S values, indicating superior overall performance. In particular, NaSrH$_{12}$ reaches an S of 1.86 at 25 GPa. These findings suggest that such compounds may achieve appreciable $T_\mathrm{c}$ at medium-to-low pressures, or even ambient pressure, and thus motivate experimental synthesis and characterization of Na$2$GaCuH$6$ at ambient pressure and cage-like hydrides (NaSrH${12}$, KSrH${12}$) at about 25 GPa using diamond anvil cells, aiming to realize high- $T_\mathrm{c}$ superconductivity at moderate pressures.

\section{Conclusion}

In summary, this work demonstrates the power of combining physics-informed feature engineering with symbolic regression to advance the understanding and discovery of hydrogen-based superconductors. The integrated DOS descriptor ($\mathrm{DOS}^{Int}$) not only addresses numerical instability in conventional calculations but also provides clear physical interpretation as the effective electron count near the Fermi surface. The model achieves  high prediction accuracy across different datasets, validating its strong generalization capability.

We successfully identified four novel hydrogen-based superconductors by using the prediction model. All four materials exhibit superconductivity indices S $>$ 1, confirming their superior overall performance.

These discoveries not only validate our physics-based interpretable machine learning approach but also establish rapid and viable new ways for achieving high-$T_\mathrm{c}$ superconductivity under milder pressure conditions. This physics-mechanism-driven data modeling framework greatly accelerates the discovery of future superconducting materials, thereby advancing the transition of hydrogen-based superconductors from theoretical prediction to practical application.

\begin{acknowledgments}
This work is supported by National Natural Science Foundation of China (Grant No. 12504069). Thanks to Dr. Xiaohua Zhang from Yanshan University for the useful discussion of effective $p$-orbital contributed from hydrogen atom. The authors thank the Center of Campus Network and Modern Educational Technology, Guangdong University of Technology, Guangdong, China, for providing computational resources and technical support for this work.
\end{acknowledgments}

\bibliographystyle{apsrev4-2}

\bibliography{cite}    
\end{document}